\documentstyle[12pt]{article}
\begin{document}
\def\bt{\begin{tabular}}
\def\et{\end{tabular}}
\def\bfr{\begin{flushright}}
\def\mm{\mbox{\boldmath $ }}
\def\efr{\end{flushright}}
\def\bfl{\begin{flushleft}}
\def\efl{\end{flushleft}}
\def\vs{\vspace}
\def\hs{\hspace}
\def\sta{\stackrel}
\def\pb{\parbox}
\def\bc{\begin{center}}
\def\ec{\end{center}}
\def\sp{\setlength{\parindent}{2\ccwd}}
\def\bp{\begin{picture}}
\def\ep{\end{picture}}
\def\uni{\unitlength=1mm}
\def\REF#1{\par\hangindent\parindent\indent\llap{#1\enspace}\ignorespaces}

\noindent \bc {\LARGE\bf Revised Iterative Solution of \\
\vspace{.2cm}Ground State of Double-Well Potential}

\vs*{5mm} {\large  Zhao Wei-Qin$^{1,~2}$}

\vs*{11mm}

{\small \it 1. China Center of Advanced Science and Technology (CCAST)}

{\small \it (World Lab.), P.O. Box 8730, Beijing 100080, China}

{\small \it 2. Institute of High Energy Physics, Chinese Academy of Sciences,

P. O. Box 918(4-1), Beijing 100039, China}

\ec
\vspace{2cm}
\begin{abstract}

A revised new iterative method based on Green function defined by
quadratures along a single trajectory is developed and applied to
solve the ground state of the double-well potential. The result is
compared to the one based on the original iterative method. The
limitation of the asymptotic expansion is also discussed.
\end{abstract}
\vspace{.5cm}

PACS{:~~11.10.Ef,~~03.65.Ge}

Key words: iterative solution, asymptotic expansion, double-well
potential

\newpage

\section*{\bf 1. Introduction}
\setcounter{section}{1}
\setcounter{equation}{0}

The double-well potential in one dimension,
\begin{eqnarray}\label{e1.1}
V(x) = \frac{1}{2} g^2 (x^2-1)^2,
\end{eqnarray}
is specially interesting since it has degenerate minima and could
be served as a simple example of the bound-state tunnelling
problem in quantum mechanics. However, it is a non-perturbative
problem to solve the Schroedinger equation for this potential due
to the tunnelling effect between the two minima. The asymptotic
series of the ground state energy is meaningful only for quite
large $g$ (e.g. $g\geq 6$). Even for such large $g$ the obtained
plateau in the energy expansion series does not necessarily
consistent to the exact solution. An effective method to obtain
the convergent solution of this problem for any values of $g$ is
needed.

Recently an iterative solution of the ground state for the one
dimensional double-well potential is obtained[1] based on the
Green function method developed in ref.[2]. This Green function is
defined along a single trajectory, from which the ground state
wave function in N-dimension can be expressed by quadratures along
the single trajectory. This makes it possible to develop an
iterative method to obtain the ground state wave function,
starting from a properly chosen trial function. The convergence of
the iterative solution very much depends on the choice of the
trial function[1].

However, in the original iterative solution the obtained
correction for the trial wave function is in the form of a power
expansion, while the bound state solution should be in the form of
an exponential when the coordinate variable approaches infinity.
Recently a new revised iterative procedure[3] based on the same
Green function is developed and applied to solve anhamornic
oscillator and Stark effect. This method has some advantages
compared to the original one. It not only gives an exponential
form for the correction of the trial wave function, but also
spends much less time in each iteration due to less folds of
integration. It is natural to try this method to solve the
double-well potential which has been attracted much attention.

In Section 2, a brief introduction is given about the Green
function method based on the single trajectory quadrature. Special
discussion is given to the revision of the iterative formula. The
revised iterative formula for the double-well potential is given
in Section 3, together with the trial function and the boundary
condition for the lowest even state. The numerical results based
on the revised iterative formula and the original iterative method
are collected in Section 4, together with a comparison to the
asymptotic expansion. A compact expression of the asymptotic
expansion is derived in Appendix A. Finally some discussions are
given at the end.

\newpage

\section*{\bf 2. Green Function and the Revised Iterative
Solution}
\setcounter{section}{2}
\setcounter{equation}{0}

\vspace{.5cm}

For a particle with unit mass, moving in an N-dimensional unperturbed
potential $V_0({\bf q})$, the ground state wave function
$\Phi({\bf q})$ satisfies the following Schroedinger equation:
\begin{eqnarray}\label{e2.1}
H\Phi({\bf q}) = E \Phi({\bf q}),
\end{eqnarray}
where
\begin{eqnarray}\label{e2.2}
H &=& T+V_0({\bf q}) = -\frac{1}{2} {\bf \nabla}^2 + V_0({\bf q}).
\end{eqnarray}
Assume the solution of eq.(\ref{e2.1}) could be expressed as
\begin{eqnarray}\label{e2.3}
\Phi({\bf q}) &=& e^{-S({\bf q})}.
\end{eqnarray}
Introduce a perturbed potential $U({\bf q})$ and assume
\begin{eqnarray}\label{e2.4}
V({\bf q})&=&V_0({\bf q})+U({\bf q})\\
{\cal H} &=& T+V({\bf q})= -\frac{1}{2} {\bf \nabla}^2 + V({\bf q}).
\end{eqnarray}
Define another wave function
$\Psi({\bf q})$ satisfying the Schroedinger equation
\begin{eqnarray}\label{e2.6}
{\cal H}\Psi({\bf q}) &=& {\cal E} \Psi({\bf q})\\
{\cal E} &=& E+\Delta.
\end{eqnarray}
Let
\begin{eqnarray}\label{e2.8}
\Psi({\bf q}) &=& e^{-S({\bf q})-\tau ({\bf q})}.
\end{eqnarray}
The equation for $\tau$
and $\Delta$ could be derived easily[2]:
\begin{eqnarray}\label{e2.9}
{\bf \nabla}S\cdot {\bf \nabla}\tau +
\frac{1}{2}[({\bf \nabla}\tau)^2 - {\bf \nabla}^2 \tau] =
(U - \Delta).
\end{eqnarray}
Consider the coordinate transformation
\begin{eqnarray}\label{e2.10}
(q_1, q_2, q_3, \cdots, q_N) \rightarrow (S, \alpha_1, \alpha_2,
\cdots, \alpha_{N-1}) = (S, \alpha)
\end{eqnarray}
with $\alpha=(\alpha_1, \alpha_2, \cdots, \alpha_{N-1})$ denoting the set
of $N-1$ orthogonal angular coordinates satisfying the condition
\begin{eqnarray}\label{e2.11}
{\bf \nabla} S\cdot {\bf \nabla} \alpha_i = 0,
\end{eqnarray}
for $i=1,~2,~\cdots~,N-1$. Some useful quantities for this
coordinate transformation[2] are listed in Appendix B. Similarly
to the discussions in Ref.[2], introducing the $\theta$-function
in S-space:
\begin{eqnarray}\label{e2.12}
\theta(S- {\overline S})=
\left\{\begin{array}{cc}
1 & ~~~~~~~~{\rm if}\hspace{4mm} 0 \leq {\overline S} < S \\
0 & ~~~~~~~~{\rm if} \hspace{4mm} 0 \leq S < {\overline S}
\end{array}
\right.
\end{eqnarray}
and define
\begin{eqnarray}\label{e2.13}
C = \theta [({\bf \nabla}S)^2]^{-1}=\theta h_S^2.
\end{eqnarray}
Using
\begin{eqnarray}\label{e2.14}
{\bf \nabla} S \cdot {\bf \nabla} C = 1,
\end{eqnarray}
it is easy to derive the following equation
\begin{eqnarray}\label{e2.15}
\tau = (1+CT)^{-1} C [(U-\Delta)-\frac{1}{2}({\bf
\nabla}\tau)^2].
\end{eqnarray}
When the N-dimensional variable ${\bf q}$ is transformed into
$(S,~\alpha)$, $T=-\frac{1}{2}{\bf \nabla}^2$ could be
decomposed into two parts:
\begin{eqnarray}\label{e2.16}
T = T_S + T_{\alpha},
\end{eqnarray}
where $ T_S$ and $T_{\alpha}$ consist only the differentiation to
$S$ and $\alpha$, respectively. The detailed expression of $ T_S$
and $T_{\alpha}$ is given in Appendix B. Now another Green
function could be defined as[2]
\begin{eqnarray}\label{e2.17}
\overline{D} &\equiv& -2\theta e^{2 S}
\frac{h_S}{h_{\alpha}} \theta e^{-2 S} h_S h_{\alpha}
\end{eqnarray}
and it is related to $C$ in the following way[2]:
\begin{eqnarray}\label{e2.18}
(1+ \overline{D}T_{\alpha})^{-1}\overline{D}=(1+CT)^{-1}C.
\end{eqnarray}
Therefore, from (\ref{e2.15}), we have
\begin{eqnarray}\label{e2.19}
\tau = (1+\overline{D}T_\alpha)^{-1} \overline{D}
[(U-\Delta)-\frac{1}{2}({\bf \nabla}\tau)^2].
\end{eqnarray}
The explicit expression of $\tau$ based on (\ref{e2.19}) is
\begin{eqnarray}\label{e2.20}
\tau =-2\int_{0}^{S}e^{2S'} \frac{h_{S' }}{h_{\alpha}}dS'
\int_{0}^{S'}e^{-2S''} h_{S''} h_{\alpha}dS''(1+T_\alpha
\overline{D})^{-1} [(U-\Delta)-\frac{1}{2}({\bf \nabla}\tau)^2].
\end{eqnarray}
Therefore, we have
\begin{eqnarray}\label{e2.21}
-\frac{1}{2} \frac{h_{\alpha} }{h_{S} } e^{-2S} \frac{\partial
\tau(S,~\alpha)}{\partial S} =\int_{0}^{S}e^{-2S'} h_{S'}
h_{\alpha}dS'(1+T_\alpha \overline{D})^{-1}
[(U-\Delta)-\frac{1}{2}({\bf \nabla}\tau)^2].
\end{eqnarray}
The left hand side of Eq.(\ref{e2.20}) approaches to $0$ when $S
\rightarrow \infty$, so is the right hand side, i.e.,
\begin{eqnarray}\label{e2.22}
\int_{0}^\infty e^{-2S} h_{S} h_{\alpha}dS(1+T_\alpha
\overline{D})^{-1} [(U-\Delta)-\frac{1}{2}({\bf \nabla}\tau)^2]=0,
\end{eqnarray}
which is correct  for all $\alpha$. Integrating over
$d\alpha=\Pi_{i=1}^{N-1}d\alpha_i$ and because of
\begin{eqnarray}\label{e2.23}
 h_{S} h_{\alpha}T_\alpha=-\frac{1}{2}
\sum_{j=1}^{N-1} \frac{\partial}{\partial \alpha_j} \frac{h_{S}
h_{\alpha}}{h_j^2}\frac{\partial}{\partial \alpha_j}
~~~{\rm and}~~~\int d \alpha h_{S} h_{\alpha}T_\alpha \tau=0,
\end{eqnarray}
we derive
\begin{eqnarray}\label{e2.24}
\int h_{S} h_{\alpha}d\alpha dS e^{-2S}
[(U-\Delta)-\frac{1}{2}({\bf \nabla}\tau)^2]=0.
\end{eqnarray}
Denoting $d {\bf q}=h_{S} h_{\alpha}d\alpha dS $, from Eq.(\ref{e2.3}),
we reach a new expression of the perturbative energy
\begin{eqnarray}\label{e2.25}
\Delta= \frac{\int d{\bf q}~\Phi^2 [U-\frac{1}{2}({\bf
\nabla}\tau)^2]}{\int d{\bf q}~\Phi^2}~.
\end{eqnarray}
Based on Eqs. (\ref{e2.25}) and (\ref{e2.15}) or (\ref{e2.19}) we have the new
iteration series
\begin{eqnarray}\label{e2.26}
\Delta_n&=&\frac{\int d{\bf q}~\Phi^2 [U-\frac{1}{2}({\bf
\nabla}\tau_{n-1})^2]}{\int d{\bf q}~\Phi^2}~,\nonumber\\
\tau_n &=& (1+\overline{D}T_\alpha)^{-1} \overline{D}
[(U-\Delta_n)-\frac{1}{2}({\bf \nabla}\tau_{n-1})^2]\\
 {\rm or}&&\nonumber\\
\tau_n &=& (1+CT)^{-1} C [(U-\Delta_n)-\frac{1}{2}
({\bf \nabla} \tau_{n-1})^2]\nonumber.
\end{eqnarray}
For later convenience this new iteration series is name as the
$\tau$-iteration in this paper. Let us compare the above
$\tau$-iteration with the original one derived from the equation
for $f=e^{-\tau}$ and $\Delta$ in Ref.[2], which is named as
$f$-iteration in this paper:
\begin{eqnarray}\label{e2.27}
\Delta_n &=& \frac{\int d{\bf q}~\Phi^2~U~f_{n-1}}
{\int d{\bf q}~\Phi^2~f_{n-1}}~\nonumber\\
f_n &=& 1+(1+\overline{D}T_\alpha)^{-1} \overline{D} (-U+\Delta_n)f_{n-1}\\
{\rm or}&&\nonumber\\
f_n &=& 1+(1+CT)^{-1} C (-U+\Delta_n)f_{n-1}.\nonumber
\end{eqnarray}
There are several advantages for the $\tau$-iteration:

1) It directly gives an exponential form for the perturbed wave
function $e^{-\tau}$. This result is consistent with those
obtained using the series expansion of $\{\bf S_i\}$ and $
\{E_i\}$ (See Section 1 of Ref.[1]);

2) The iteration process in this formula is more transparent;

3) The calculation of the perturbation energy is much simpler.

\vspace{.5cm}

\newpage

\section*{\bf 3. Revised Iterative Formula for the Double-well \\
Potential}
\setcounter{section}{3}
\setcounter{equation}{0}

\vspace{.5cm}

In this section the revised iterative formula is applied to solve
the ground state for the double-well potential. For the
Hamiltonian $H=T+V$ let us introduce the wave function for the
lowest even eigenstate as $\psi_{ev}$, satisfying
\begin{eqnarray}\label{e3.1}
H \psi_{ev} &=& E_{ev} \psi_{ev}.
\end{eqnarray}
In the following we are going to introduce the trial wave function
$\phi_{ev}$ for this state, satisfying
\begin{eqnarray}\label{e3.2}
(H + w_{ev}) \phi_{ev} &=& (E_{ev} +{\cal E}_{ev}) \phi_{ev}=g
\phi_{ev}
\end{eqnarray}
Assuming
\begin{eqnarray}\label{e3.3}
\psi_{ev} &=& \phi_{ev} e^{-\tau_{ev}},
\end{eqnarray}
the final energy and wave function $\{\psi_{ev},~E_{ev}\}$ could
be obtained by solving the corresponding equation of
$\{\tau_{ev},~{\cal E}_{ev}\}$ based on the revised iteration
method introduced in Section 2. The key to choose the proper trial
function is to satisfy the necessary boundary conditions of the
state. For the ground state, namely the lowest even state, we have
\begin{eqnarray}\label{e3.4}
\psi_{ev}(-x) &=& \psi_{ev}(x)\nonumber\\
\psi'_{ev}(0) &=& 0\\
\psi_{ev}(\infty) &=& 0.\nonumber
\end{eqnarray}
The trial wave function should satisfy similar boundary
conditions, namely,
\begin{eqnarray}\label{e3.5}
\phi_{ev}(-x) &=& \phi_{ev}(x)\nonumber\\
\phi'_{ev}(0) &=& 0\\
\phi_{ev}(\infty) &=& 0.\nonumber
\end{eqnarray}

\newpage

Following the steps in Section 2 of Ref.[1], we introduce, for $x
\geq 0$,
\begin{eqnarray}\label{e3.6}
g S_0(x) & \equiv & \frac{g}{3}(x-1)^2(x+2),\nonumber\\
S_1(x) &\equiv & ln~\frac{x+1}{2},\\
\phi_+(x)  = \phi_+(-x) &\equiv&  e^{-g S_0(x)-S_1(x)} =  e^{-g S_0(x)}(\frac{2}{1+x}),
\nonumber\\
\phi_-(x) &\equiv &  e^{-gS_0(-x)-S_1(x)}=
e^{-\frac{4}{3}g}e^{+gS_0(x)}(\frac{2}{1+x}).\nonumber
\end{eqnarray}
It is easy to see that $\phi_+(x)$ satisfies
\begin{eqnarray}\label{e3.7}
(T+V+u)\phi_+=g\phi_+,
\end{eqnarray}
where
\begin{eqnarray}\label{e3.8}
V(x) &=& \frac{1}{2}g^2 (x^2-1)^2\nonumber\\
u(x) &=& \frac{1}{(1+x)^2}.
\end{eqnarray}
Following the necessary boundary conditions (\ref{e3.5}), we could
choose the trial function as
\begin{eqnarray}\label{e3.9}
\phi_{ev}(x) = \phi_{ev}(-x) \equiv
\left\{\begin{array}{ccc}
\phi_+(x) + \frac{g-1}{g+1} \phi_-(x),~~~~~~~{\sf for}~~0 \leq x <1\\
(1+\frac{g-1}{g+1}e^{-\frac{4}{3}g})\phi_+(x),~~~~~~{\sf for}~~x>1.
\end{array}
\right.
\end{eqnarray}
It is easy to proof that the above trial function satisfies Eq.(\ref{e3.2}) with
\begin{eqnarray}\label{e3.10}
w_{ev}(x) &=& w_{ev}(-x)=u(x) + \hat{g}_{ev} (x),~~~~~~ {\sf for}~~~ x \geq 0\\
\hat{g}_{ev} (x) &=&
\left\{\begin{array}{ll}
2g\frac{(g-1) e^{2g S_0(x)-\frac{4}{3} g}}{(g+1)+(g-1)e^{2g S_0(x)-\frac{4}{3} g}},
&~~~~~~{\sf for}~~0 \leq x<1\\
0&~~~~~~{\sf for}~~x>1.
\end{array}\nonumber
\right.
\end{eqnarray}
Although the potentials $V+w_{ev}$ is not continuous at $x=1$, it
could be proved that  $\phi_{ev}$ and $\phi'_{ev}$ are continuous
at any $x$, including $x=1$ and $x=0$.

Now introduce the $\theta$-function in $x$-space:
\begin{eqnarray}\label{e3.11}
(x|\theta|y)=
\left\{\begin{array}{ccc}
0~~~~~&{\sf for}~~~~~&x>y,\\
-1~~~~~&{\sf for}~~~~~&x<y
\end{array}
\right.
\end{eqnarray}
and define the Green function
\begin{eqnarray}\label{e3.12}
\overline{D}_{ev}~ &\equiv& -2 \theta
\phi_{ev}^{-2}\theta\phi_{ev}^2.
\end{eqnarray}
Following similar steps as in Section 2, we obtain the equations
for $\{\tau_{ev},~{\cal E}_{ev}\}$ as follows:
\begin{eqnarray}\label{e3.13}
\tau_{ev} &=& \overline{D}_{ev}[-(w_{ev} - {\cal E}_{ev})-\frac{1}{2}(\tau'_{ev})^2],\\
{\cal E}_{ev} &=& \frac{\int\limits_0^{\infty}\phi^2_{ev}~
(w_{ev}+\frac{1}{2}~(\tau'_{ev})^2) dx} {\int\limits_0^{\infty}
\phi^2_{ev} dx}.\nonumber
\end{eqnarray}
The solution of the ground state is
\begin{eqnarray}\label{e3.14}
\psi_{ev} &=& \phi_{ev}~ e^{-\tau_{ev}},~~~~~E_{ev}=g-{\cal
E}_{ev}.
\end{eqnarray}
Similar procedure could also be taken by introducing
\begin{eqnarray}\label{e3.15}
\overline{D}_{+}~ &\equiv& -2 \theta \phi_{+}^{-2}\theta\phi_{+}^2
\end{eqnarray}
and
\begin{eqnarray}\label{e3.16}
\tau_{+} &=& \overline{D}_{+}[-(u - {\cal E}_{+})-\frac{1}{2}(\tau'_{+})^2],\\
{\cal E}_{+} &=& \frac{\int\limits_0^{\infty}\phi^2_{+}~ (u+
\frac{1}{2}~(\tau'_{+})^2) dx} {\int\limits_0^{\infty} \phi^2_{+}~
 dx}.\nonumber
\end{eqnarray}
It should be noticed that although $\psi_+= \phi_+~ e^{-\tau_+}$
does satisfy the equation
\begin{eqnarray}\label{e3.17}
H\psi_+=E_+\psi_+,
\end{eqnarray}
with $E_+ = g - {\cal E}_{+}$, the solution $\psi_+$ is not the
eigenstate of the Hamiltonian $H=T+V$, because the trial function
$\phi_+$, as well as the solution $\psi_+$ do not satisfy the
necessary boundary conditions. However, we still like to keep this
solution here since an analytic expression of $E_+$ and $\psi_+$
has been obtained in terms of the asymptotic power expansion of
$1/g$ in Ref.[1]. This makes it possible to check the accuracy of
the iteration procedure. In Appendix A compact expressions of
the asymptotic power expansion of $E_+$ and $\tau'_+$ are derived.

Based on the explicit integral expression of $\overline{D}_{ev}$
and $\overline{D}_+$, taking derivatives of $\tau_{ev}$ and
$\tau_+$ in the first equations of (\ref{e3.13}) and
(\ref{e3.16}), we obtain
\begin{eqnarray}\label{e3.18}
\tau'_{ev}(x) &=& 2 \phi^{-2}_{ev}(x)\int\limits_0^{x}\phi^2_{ev}(y)
          [(w_{ev}(y) - {\cal E}_{ev})+\frac{1}{2}(\tau'_{ev}(y))^2]~dy,\\
{\cal E}_{ev} &=& \frac{\int\limits_0^{\infty}\phi^2_{ev}(x)~
(w_{ev}(x)+\frac{1}{2}~(\tau'_{ev}(x))^2) dx} {\int\limits_0^{\infty}
\phi^2_{ev}(x)~ dx}.\nonumber
\end{eqnarray}
and
\begin{eqnarray}\label{e3.19}
\tau'_{+}(x) &=& 2 \phi^{-2}_{ev}(x) \int\limits_0^{x}\phi^2_{ev}(y)
          [(u(y) - {\cal E}_{+})+\frac{1}{2}(\tau'_{+}(y))^2]~dy,\\
{\cal E}_+ &=& \frac{\int\limits_0^{\infty}\phi^2_{+}(x)~
(u(x)+\frac{1}{2}~(\tau'_{+}(x))^2) dx} {\int\limits_0^{\infty}
\phi^2_{+}(x)~ dx}.\nonumber
\end{eqnarray}
The two sets of equations (\ref{e3.18}) and (\ref{e3.19}) are for
the pairs of $\tau'_{ev},~{\cal E}_{ev}$ and $\tau_+,~{\cal E}_+$,
which could be solved iteratively in the following way:
Introducing the initial conditions
\begin{eqnarray}\label{e3.20}
{\cal E}_{ev,0} &=& 0,~~~~~~~~~~\tau'_{ev,0}=0
\end{eqnarray}
and
\begin{eqnarray}\label{e3.21}
{\cal E}_{+,0} &=& 0,~~~~~~~~~~\tau'_{+,0}=0,
\end{eqnarray}
we have, for $n=1$,
\begin{eqnarray}\label{e3.22}
{\cal E}_{ev,1} &=&
      \frac{\int\limits_0^{\infty}\phi^2_{ev}(x)~w_{ev}(x) dx}
                    {\int\limits_0^{\infty} \phi^2_{ev}(x)dx}\\
\tau'_{ev,1}(x) &=& 2 \phi^{-2}_{ev}(x) \int\limits_0^{x}\phi^2_{ev}(y)
                (w_{ev}(y) - {\cal E}_{ev,1})dy\nonumber\\
               &=& -2 \phi^{-2}_{ev}(x) \int\limits_{x}^{\infty}\phi^2_{ev}(y)
                (w_{ev}(y) - {\cal E}_{ev,1})dy\nonumber
\end{eqnarray}
and
\begin{eqnarray}\label{e3.23}
{\cal E}_{+,1} &=& \frac{\int\limits_0^{\infty}\phi^2_+(x)~u(x) dx}
                    {\int\limits_0^{\infty} \phi^2_+(x) dx}.\\
\tau'_{+,1}(x)  &=& 2 \phi^{-2}_+(x)\int\limits_0^{x}\phi^2_+(y)(u(y)
                - {\cal E}_{+,1}) dy\nonumber\\
            &=& -2 \phi^{-2}_+(x) \int\limits_{x}^{\infty}\phi^2_+(y)(u(y)
               - {\cal E}_{+,1}) dy\nonumber
\end{eqnarray}
For $n>1$ we have

\newpage

\begin{eqnarray}\label{e3.24}
{\cal E}_{ev,n} &=& {\cal E}_{ev,1}+\frac{\int\limits_0^{\infty}\phi^2_{ev}(x)~
               \frac{1}{2}(\tau'_{ev,n-1}(x))^2 dx} {\int\limits_0^{\infty}
                 \phi^2_{ev}(x) dx}\\
\tau'_{ev,n}(x) &=&  \tau'_{ev,1}(x)+2 \phi^{-2}_{ev}(x) \int\limits_0^{x}\phi^2_{ev}(y)
                  [({\cal E}_{ev,1} - {\cal
         E}_{ev,n})+\frac{1}{2}(\tau'_{ev,n-1}(y))^2]dy~~~~~~\nonumber\\
           &=&  \tau'_{ev,1}(x)-2 \phi^{-2}_{ev}(x) \int\limits_{x}^{\infty}\phi^2_{ev}(y)
                  [({\cal E}_{ev,1} - {\cal E}_{ev,n})
                  +\frac{1}{2}(\tau'_{ev,n-1}(y))^2]dy\nonumber
\end{eqnarray}
and
\begin{eqnarray}\label{e3.25}
{\cal E}_{+,n} &=& {\cal E}_{+,1}+\frac{\int\limits_0^{\infty}\phi^2_{+}(x)~
              \frac{1}{2}(\tau'_{+,n-1}(x))^2 dx} {\int\limits_0^{\infty}
              \phi^2_{+}(x)~ dx}.\\
\tau'_{+,n}(x)  &=&  \tau'_{+,1}(x)+2 \phi^{-2}_+(x)
             \int\limits_0^{x}\phi^2_+(y)[({\cal E}_{+,1} - {\cal E}_{+,n})
               +\frac{1}{2}(\tau'_{+,n-1}(y))^2]dy\nonumber\\
            &=&  \tau'_{+,1}(x)-2 \phi^{-2}_+(x)
             \int\limits_{x}^{\infty}\phi^2_+(y)[({\cal E}_{+,1} - {\cal E}_{+,n})
               +\frac{1}{2}(\tau'_{+,n-1}(y))^2]dy\nonumber
\end{eqnarray}
For comparison we list in the following also the $f$-iteration
formula for the double-well potential based on (\ref{e2.27})[2]:
\begin{eqnarray}\label{e3.26}
f_n(x) &=& 1 -2\int\limits_x^{\infty} \phi^{-2}(y) dy
\int\limits_y^{\infty} \phi^2(z) (w(z) -  {\cal E}_n)f_{n-1}(z)
dz\\
{\cal E}_{n} &=& \int\limits_0^{\infty} \phi^2(x)f_{n-1}(x) w(x)
dx \bigg/ \int\limits_0^{\infty}\phi^2(x) f_{n-1}(x) dx,\nonumber
\end{eqnarray}
where $f_n$ and ${\cal E}_n$ could be either for the even ground
state "$ev$" or the "$+$" state.

In the following section we are going to show the iterative
results of ${\cal E}_{+,n}$ and ${\cal E}_{ev,n}$ based on the
$\tau$-iteration formula Eqs.(\ref{e3.24}) and (\ref{e3.25}),
together with the result from the $f$-iterative formula
(\ref{e3.26}). The obtained ${\cal E}_{+,n}$ is also compared to
the asymptotic expansion for different $g$, obtained from the
compact expression in Appendix A.

\newpage

\section*{\bf 4. Numerical Results and Discussions}
\setcounter{section}{5}
\setcounter{equation}{0}

\vspace{.5cm}

Now let us look at the results based on the $\tau$-iteration
formula (\ref{e3.24}) and (\ref{e3.25}). Starting from the trial
wave functions $\phi_{ev}$ and $\phi_{+}$ defined in (\ref{e3.9})
and (\ref{e3.6}), the energies ${\cal E}_{ev}$ and ${\cal E}_+$
after the first 4 steps of iteration with 5-fold integrations are
listed in Tables 1 and 2, together with the energies
$E_{ev,5}=g-{\cal E}_{ev,5}$ and $E_{+,5}=g-{\cal E}_{+,5}$. In
Ref.[1] a convergent iteration method to solve the lowest states
of the double-well potential, i.e. the $f$-iteration, has been
given. However, no numerical results have been provided. For
comparison the numerical results for ${\cal E}_{ev}$ and ${\cal
E}_+$ based on the $f$-iteration method is also listed in Tables 1
and 2. For the same folds of integrations the solution of the
$f$-iteration could only reach a lower accuracy, comparing to the
$\tau$-iteration, because each step iteration of $\tau'$ needs
only one fold of integration, while each step iteration of $f$
depends on 2-fold integrations. The accuracy of the obtained
energies is higher when $g$ becomes larger.

\vspace{1cm}

 Table 1. ${\cal E}_{ev,n}$ and $E_{ev,5}=g-{\cal E}_{ev,5}$

\begin{tabular}{|c|c|c|c|c|c|c|c|}
  \hline
  $g$ & $n$ & 1 & 2 & 3 & 4 & 5 & $E_{ev,5}$\\
  \hline
  0.05 & $\tau$-iter. & -0.0341 & -0.0172 & -0.0158 & -0.0163 & -0.0164 & 0.0664\\
  \hline
  0.1 & $\tau$-iter. & -0.0118 & -0.0022 & -0.0016 & -0.0017 & -0.0017 & 0.1017\\
  \hline
  0.3 & $\tau$-iter. & 0.0963 & 0.0973 & 0.0973 & 0.0973   & 0.0973 & 0.2027\\
  \hline
  0.5 & $\tau$-iter. & 0.2035 & 0.2060 & 0.2060 & 0.2060 & 0.2060 & 0.2940\\
  \hline
  0.5 & $f$-iter. & 0.2036 & 0.2056 & 0.2060 &    &&\\
  \hline
  1 & $\tau$-iter. & 0.4135 & 0.4310 & 0.4312 & 0.4311 & 0.4311 & 0.5689\\
  \hline
  1 & $f$-iter. & 0.4135 & 0.4267 & 0.4302 &    &&\\
  \hline
  3 & $\tau$-iter. & 0.4757 & 0.5105 & 0.5166 & 0.5173 & 0.5173 & 2.4827 \\
  \hline
  3 & $f$-iter. & 0.4757 & 0.5053 & 0.5141 &   &&\\
  \hline
  6 & $\tau$-iter. & 0.29204 & 0.29399 & 0.29420 & 0.29422 & 0.29422 & 5.70578\\
  \hline
  6 & $f$-iter. & 0.29204 & 0.29393 & 0.29419   & &&\\
  \hline
  7 & $\tau$-iter. & 0.27884 & 0.27957 & 0.27962 & 0.27963 & 0.27963 & 6.72037\\
  \hline
  7 & $f$-iter. & 0.27884 & 0.27955 & 0.27961   & &&\\
  \hline
  8 & $\tau$-iter. & 0.27231 & 0.27265 & 0.27266 & 0.27266 & 0.27266 & 7.72734 \\
  \hline
  8 & $f$-iter. & 0.27231 & 0.27264 & 0.27266   & &&\\
  \hline
\end{tabular}

\newpage

Table 2. ${\cal E}_{+,n}$ and $E_{+,5}=g-{\cal E}_{+,5}$

\begin{tabular}{|c|c|c|c|c|c|c|c|}
  \hline
  $g$ & $n$ & 1 & 2 & 3 & 4 & 5 & $E_{+,5}$\\
  \hline
  1 & $\tau$-iter. & 0.4135 & 0.4310 & 0.4312 & 0.4311 & 0.4311 & 0.5689\\
  \hline
  1 & $f$-iter. & 0.4135 & 0.4267 & 0.4302 &  &  &\\
  \hline
  3 & $\tau$-iter. & 0.3221 & 0.3257 & 0.3258 & 0.3258 & 0.3258 & 2.6742\\
  \hline
  3 & $f$-iter. & 0.3221& 0.3254 & 0.3257 &  & &\\
  \hline
  6 & $\tau$-iter. & 0.27989 & 0.28040 & 0.28041 & 0.28041 & 0.28041 & 5.71959 \\
  \hline
  6 & $f$-iter. & 0.27989 & 0.28039 & 0.28041   & &&\\
  \hline
  7 & $\tau$-iter. & 0.27461 & 0.27494 & 0.27494 & 0.27494 & 0.27494 & 6.72506 \\
  \hline
  7 & $f$-iter. & 0.27461 & 0.27493 & 0.27494   & &&\\
  \hline
  8 & $\tau$-iter. & 0.27091 & 0.27113 & 0.27113 & 0.27113 & 0.27113 & 7.72887 \\
  \hline
  8 & $f$-iter. & 0.27091 & 0.27113 & 0.27113   & &&\\
  \hline
\end{tabular}

\vspace{1cm}

It is shown clearly that the obtained energies $E_{ev}$ and $E_+$
are lower than $g$ and $E_{ev}<E_+$, which are reasonable. When
$g$ increases the two energies become very close to each other and
the second step of iteration gives already quite accurate result.
It is interesting to notice that for $g=1$ the trial function
$\phi_+$ and $\phi_{ev}$ are the same, therefore starting from
$\phi_+$ the obtained $\psi_+$ by iteration is the exact ground
state wave function for $g=1$.

\vspace{1cm}

Table 3. ${\cal E}_{ev}$ and $E_{ev}=g-{\cal E}_{ev}$

\begin{tabular}{|c|c|c||c|c|c|}
  \hline
  $g$ & ${\cal E}_{ev}$ & $E_{ev}$ & $g$ & ${\cal E}_{ev}$ & $E_{ev}$  \\
  \hline
  0.05 & -0.0164 & 0.0664 & 2.2 & 0.5951 & 1.6049  \\
  \hline
  0.1 & -0.0017 & 0.1017 & 2.5 & 0.5745 & 1.9255 \\
  \hline
  0.3 & 0.0973 & 0.2027 & 2.7 & 0.5539 & 2.1461 \\
  \hline
  0.5 & 0.2060 & 0.2940 &  3.0 & 0.5173 & 2.4827  \\
  \hline
  0.7 & 0.3065 & 0.3935 & 4.0 & 0.3984 & 3.6016  \\
  \hline
  1.0 & 0.4311 & 0.5689 & 5.0 & 0.3273 & 4.6727 \\
  \hline
  1.5 & 0.5598 & 0.9402 & 6.0 & 0.29422 & 5.70578  \\
  \hline
  1.7 & 0.5851 & 1.1149 & 7.0 & 0.27963 & 6.72037  \\
  \hline
  2.0 & 0.5990 & 1.4010 & 8.0 & 0.27266 & 7.72734  \\
  \hline
\end{tabular}

\vspace{1cm}

In Table 3 listed are the obtained ${\cal E}_{ev}$ for different
$g$ based on the $\tau$-iteration. The obtained ${\cal E}_{ev}$ is
negative for very small $g$ and increases when $g$ increases from
$0.05$ to $2.0$, then decreases when $g$ increases further.
However, the energy for the ground state $E_{ev}=g-{\cal E}_{ev}$
monotonically increases with increasing $g$.

\newpage

Table 4. ${\cal E}_{+}$, $E_+=g-{\cal E}_{+}$ and ${\cal
E}_{+}^N=\sum\limits_{m=0}^N \epsilon_{m+1}/g^m$

\begin{tabular}{|c|c|c|c|c|c|c|}
  \hline
  $g$ & ${\cal E}_+$ & $E_+$ & $N_{min}$ & $N_{max}$ & ${\cal E}_+^N$ & $O(e^{-\frac{4}{3}g})$  \\
  \hline
  1 & 0.4311 & 0.5689 & - & - & - &  $\sim$ 0.3\\
  \hline
  2 & 0.3664 & 1.6336 & - & - & -  & $\sim$ 0.07  \\
  \hline
  3 & 0.3258 & 2.6742 & - & - & - &  $\sim$ 0.02    \\
  \hline
  4 & 0.3024 & 3.6976 & - & - & -  & $\sim$ 0.005  \\
  \hline
  5 & 0.2888 & 4.7112 & - & - & - & $\sim$ 0.001   \\
  \hline
  6 & 0.28041 & 5.71959 & 10 & 17 & 0.2807 & $\sim~ 3\times 10^{-4}$    \\
  \hline
  7 & 0.27494 & 6.72506 & 14 & 22 & 0.27501 & $\sim~ 4\times 10^{-5}$   \\
  \hline
  8 & 0.27113 & 7.72887 & 12 & 34 & 0.27115 &   $\sim~ 2\times 10^{-5}$ \\
  \hline
  9 & 0.268336 & 8.731664 & 11 & 35 & 0.268339 & $\sim 6\times 10^{-6}$  \\
  \hline
\end{tabular}

\vspace{1cm}

The energy ${\cal E}_+$ and $E_+$ are given in Table 4 together
with $\epsilon_+$ calculated from the asymptotic power expansion
to $1/g$. It can be seen that only when $g$ is large enough, say
$g\geq 6$, the asymptotic expansion is meaningful. For such large
$g$ the obtained ${\cal E}_+$ from the iteration and from the
power expansion are comparable up to a quite accurate level. The
power expansion of ${\cal E}_+$ to $1/g$ is an asymptotic one. For
a fixed and large enough $g$-value the summation up to a certain
number of terms becomes stable. When increasing the number of
summed terms further a plateau of the energy ${\cal E}_+^N$
(within the number of summed terms $N_{min}<N<N_{man}$ shown in
Table 4) is obtained, which gives the $E_+$-value accurate to a
certain level. However, beyond a certain number of terms
($N>N_{max}$) the result becomes unstable and meaningless. It
should be noticed that the asymptotic expansion result within the
plateau region does not give the accurate value of the energy. It
differs from the iteration one in the order of
$e^{-\frac{4}{3}g}$, which provides the limitation of the accuracy
of the asymptotic expansion.

Recently it has been proved[5] that the $f$-iteration in
one-dimensional problem is convergent if the trial function is
properly chosen to have a finite perturbed potential $w(x)$,
satisfying the conditions $w(x)>0$, $w'(x)<0$ and $w(\infty)=0$.
There is no restriction to the magnitude of $w(x)$. For the
double-well potential, $w_{ev}(x)$ defined in (\ref{e3.10})
satisfies these conditions when $g\geq 1$. Our numerical results
show that both $f$- and $\tau$-iteration are convergent for $g\geq
1$, although it is not an easy task to prove the convergence of
the $\tau$-iteration. Furthermore, the two iteration series could
also be applied in some region of $g<1$, where $w_{ev}(x)$ is
still positive and finite but no more a monotonic function of $x$,
and reasonable results are obtained. However, for very small $g$
(e.g. $g<0.4$), where $w(x)$ becomes negative in some $x$-region,
the $f$-iteration does not give reasonable results and the
obtained ${\cal E}_n$ becomes unstable, while the $\tau$-iteration
could still work, although the convergence becomes slower when $g$
is smaller. The reason is the following: The iterative formula for
the perturbed wave function $f$ is expressed as sum of two terms.
It could become negative if the term containing $w(x)$ becomes
negative. This would give a negative $\psi(x)$ in some $x$-region,
while the solution for the ground state wave function $\psi(x)$
should be always positive. For the $\tau$-iteration the perturbed
wave function $e^{-\tau(x)}$ is always positive for any finite
$\tau(x)$, either positive or negative. This condition is
fulfilled as long as $|w(x)|$ is finite and $w(\infty) \rightarrow
0$. Thus, the $\tau$-iteration gives less restrictions to the
perturbed potential $w(x)$. Therefore, it is of interests to
further study the condition for the convergence of the
$\tau$-iteration and to apply it to other physics problems where
perturbation method could not be applied.

\section*{\bf Acknowledgment}

The author is grateful to Professor T. D. Lee for his continuous
and substantial instructions and advice. This work is partly
supported by National Natural Science Foundation of China (NNSFC,
No. 20047001).


\bc {\Large \bf Reference} \ec

1. R. Friedberg, T. D. Lee, W. Q. Zhao and A. Cimenser, Ann. Phys.
294(2001)67

2. R. Friedberg, T. D. Lee, W. Q. Zhao, Ann. Phys. 288(2001)52

3. Zhao Wei-Qin, Commun. Theoret. Phys. (Beijing,China) 42(2004)37

4. R. Friedberg, T. D. Lee and W. Q. Zhao, IL Nuovo Cimento
A112(1999)1195

5. R. Friedberg and T. D. Lee, Ann. Phys. 308(2003)263, quant-ph/0407207

\newpage

\section*{\bf Appendix }
\setcounter{section}{10}
\setcounter{equation}{0}

\noindent
{\bf Appendix A. Asymptotic series expansion of $\tau_+$ and ${\cal E}_+$ }\\

Starting from the integral equation of $\tau'_+$ of the first equation
in (\ref{e3.20})
$$
\tau_+'(x)= 2 \phi_+^{-2}(x) \int\limits_x^{\infty}
\phi_+^2(y)[-(u(y)-{\cal E}_+)-\frac{1}{2}~(\tau_+'(y))^2] dy
\eqno(A.1)
$$
an asymptotic series expansion of $\tau_+$ and
${\cal E}_+$ could be obtained. From (A.1)
it is easy to obtain the following equation
$$
(\frac{1}{2}~\phi_+^2\tau'_+)' = [(u-{\cal E}_+)+
\frac{1}{2}~(\tau'_+)^2]\phi_+^2~ \eqno(A.2)
$$
Using the definition of $\phi_+$ in (\ref{e3.6}), the above equation
leads to
$$
g S_0'\tau'_+= \frac{1}{2}~\tau_+'' - S_1'\tau'_+ - (u-{\cal E}_+)
-\frac{1}{2}~(\tau'_+)^2. \eqno(A.3)
$$
Now let us expand both $\tau'_+(x)$ and ${\cal E}_+$ in power
series of $g^{-1}$ as following
$$
\tau_+'= \sum\limits^{\infty}_1 \frac{1}{g^m}~ S_{m+1}' ~~~{\rm
and}~~~  {\cal E}_+ = \sum\limits^{\infty}_0 \frac{1}{g^m}~
\epsilon_{m+1}.
\eqno(A.4)
$$
Substituting (A.4) into (A.3),
comparing terms proportional to $g^{-n}$, we obtain a series of
equations for $\{S_m\}$ and $\{\epsilon_m\}$:
\begin{eqnarray}
g^{0}~~~~~~~~~~S_0'S_2' &=& -(u-\epsilon_1)\nonumber\\
g^{-1}~~~~~~~~~~S_0'S_3' &=& \frac{1}{2}~S_2''-S_1'S_2'+\epsilon_2\nonumber\\
g^{-2}~~~~~~~~~~S_0'S_4' &=&
\frac{1}{2}~S_3''-S_1'S_3'+\epsilon_3-\frac{1}{2}~S^{\prime 2}_2
~~~~~~~~~~~~~~~~~~~~~~~~~~~~~~~~~~~~~~~~~~~~~~(A.5)\nonumber\\
&&\cdots\nonumber\\
g^{-m}~~~~~~S_0'S_{m+2}' &=&
\frac{1}{2}~S_{m+1}''-S_1'S_{m+1}'+\epsilon_{m+1}
-\frac{1}{2}~\sum\limits_{n=1}^{m-1}S'_{n+1}S'_{m+1-n}.\nonumber
\end{eqnarray}
The above equations are exactly the same as those obtained in
Ref.[5], when taking ${\cal E}_+=-E$ and $\epsilon_m=-E_m$. From
the first equation of (A.5), considering $u=\frac{1}{(1+x)^2}$, we
have $\epsilon_1=\frac{1}{4}$ when setting $x=1$.

In the following we derive a compact expression of $\{S'_m\}$ and
$\{\epsilon_m\}$. Assuming, for $m\geq 1$,
$$
S_{m+1}' = \frac{1}{2^{4m}}~\xi^2 \sum\limits_{i=0}^{2m-1}
\beta_i(m)\xi^i \eqno(A.6)
$$
where
$$
\xi=\frac{2}{1+x}. \eqno(A.7)
$$
Considering $S_1'=\frac{1}{x+1}$, from the last equation of (A.5)
we have
$$
S_0'S_{m+2}'=  - \frac{1}{2^{4m+2}}~\sum\limits^{2m-1}_{l=0}
\beta_l(m)\xi^{l+3}(l+4)~~~~~~~~~~~~~~~~~~~~~~~~~~~~~~~~~~~~~~~~~~~~~~
$$
$$
-\frac{1}{2^{4m+2}}~\sum\limits^{m-1}_{n=1}\sum\limits^{2n-1}_{i=0}
\sum\limits^{2(m-n)-1}_{j=0}2\beta_i(n)\beta_j(m-n)\xi^{i+j+4}+\epsilon_{m+1}.
\eqno(A.8)
$$
For the second summation on the right hand side of (A.8), defining
$i+j+1=l$, we have $l_{min}=1$ and $l_{max}=2m-1$, which leads to
$$
S_0'S_{m+2}'=  - \frac{1}{2^{4m+2}}~\sum\limits^{2m-1}_{l=0}
\beta_l(m)\xi^{l+3}(l+4)~~~~~~~~~~~~~~~~~~~~~~~~~~~~~~~~~~~~~~~~~~~~~~
$$
$$
-\frac{1}{2^{4m+2}}~\sum\limits^{2m-1}_{l=0}\sum\limits^{m-1}_{n=1}
\sum\limits^{i_{max}}_{i=i_{min}}2\beta_i(n)\beta_{l-i-1}(m-n)\xi^{l+3}+\epsilon_{m+1},
\eqno(A.9)
$$
where
$$
i_{min}={\rm max}(0,l-2(m-n))
$$
$$
i_{max}={\rm min}(2n-1,l-1).\eqno(A.10)
$$
To fix $\epsilon_{m+1}$ we put $x=1$, then $\xi=1$ and
$S_0'=1-x^2=0$. This gives
$$
\epsilon_{m+1}= \frac{1}{2^{4m+2}}~\sum\limits^{2m-1}_{l=0}
\beta_l(m)(l+4)+\frac{1}{2^{4m+2}}~\sum\limits^{2m-1}_{l=1}\sum\limits^{m-1}_{n=1}
\sum\limits^{i_{max}}_{i=i_{min}}2\beta_i(n)\beta_{l-i-1}(m-n).
\eqno(A.11)
$$
Introducing $S_0'=-\frac{4(\xi-1)}{\xi^2}$, $S_{m+2}'$ could be
expressed as


$$
S_{m+2}'=
\frac{1}{2^{4(m+1)}}~\xi^2~\frac{1}{\xi-1}~\{\sum\limits^{2m-1}_{l=0}
\beta_l(m)(l+4)(\xi^{l+3}-1)~~~~~~~~~~~~~~~
$$
$$
+\sum\limits^{2m-1}_{l=1}\sum\limits^{m-1}_{n=1}
\sum\limits^{i_{max}}_{i=i_{min}}2\beta_i(n)\beta_{l-i-1}(m-n)(\xi^{l+3}-1)\}.
\eqno(A.12)
$$
Applying the equality
$$
\xi^{l+3}-1=(\xi-1)(\xi^{l+2}+\xi^{l+1}+\cdots+1)=(\xi-1)\sum\limits_{L=0}^{l+2}\xi^L,
\eqno(A.13)
$$
changing the summation order of $L$ and $l$ in (A.12) we finally
reach the following expression of $S_{m+2}'$:

\newpage

$$
S_{m+2}'=\frac{1}{2^{4(m+1)}}~\xi^2~\sum\limits_{L=0}^{2m+1}~\xi^L\{\sum\limits^{2m-1}_{l=l_1}
\beta_l(m)(l+4)+\sum\limits^{2m-1}_{l=l_2}\sum\limits^{m-1}_{n=1}
\sum\limits^{i_{max}}_{i=i_{min}}2\beta_i(n)\beta_{l-i-1}(m-n)\}
$$
$$
=\frac{1}{2^{4(m+1)}}\xi^2~\sum\limits_{L=0}^{2m+1}\beta_L(m+1)\xi^L
~~~~~~~~~~~~~~~~~~~~~~~~~~~~~~~~~~~~~~~~~~~~~ \eqno(A.14)
$$
with
$$
l_1={\rm max}(0,L-2),~~~~~~l_2={\rm max}(1,L-2).
$$
For $m=1$, it is easy to get $\beta_0(1)=1$ and $\beta_1(1)=1$.
Finally we obtain , for $m\geq 1$,
$$
\beta_L(m+1)=\beta_l^0(m+1)+\Delta\beta_l(m+1) \eqno(A.15)
$$
where
$$
\beta_L^0(m+1)=\sum\limits^{2m-1}_{l={\rm max}(0,L-2)}
\beta_l(m)(l+4)~~~~~~~~~~~~~~~~~~ \eqno(A.16)
$$
$$
\Delta\beta_L(m+1)=\sum\limits^{2m-1}_{l={\rm
max}(1,L-2)}\sum\limits^{m-1}_{n=1}
\sum\limits^{i_{max}}_{i=i_{min}}2\beta_i(n)\beta_{l-i-1}(m-n)
$$
Taking $L=0$ in (A.15) and (A.16), from (A.11) we have
$$
\epsilon_{m+1}=\frac{1}{2^{4m+2}}~\beta_0(m+1). \eqno(A.17)
$$
Based on (A.15) and (A.16) a revised pyramid structure of
$\beta_l(m)$ could be constructed in a similar way as those  in
Appendix D of Ref.[1].
\begin{eqnarray*}
\begin{array}{ccccccl}
&&\beta_1(1)&\beta_0(1)&&&~~~m=1\\
&&&&&&\\
&\beta^0_3(2)&\beta^0_2(2)&\beta^0_1(2)&\beta^0_0(2)&&~~~\\
&\Delta\beta_3(2)&\Delta\beta_2(2)&\Delta\beta_1(2)&\Delta\beta_0(2)&&~~~\\
&\beta_3(2)&\beta_2(2)&\beta_1(2)&\beta_0(2)&&~~~m=2\\
&&&&&&\\
~~~~~~~~\beta^0_5(3)&\beta^0_4(3)&\beta^0_3(3)&\beta^0_2(3)&\beta^0_1(3)&\beta^0_0(3)&~~~\\
~~~~~~~~\Delta\beta_5(3)&\Delta\beta_4(3)&\Delta\beta_3(3)&\Delta\beta_2(3)&\Delta\beta_1(3)&\Delta\beta_0(3)&~~~\\
~~~~~~~~\beta_5(3)&\beta_4(3)&\beta_3(3)&\beta_2(3)&\beta_1(3)&\beta_0(3)&~~~m=3
~~~~~~~~~~~~~~~~~(A.18)\\
&&\cdots&\cdots&&
\end{array}
\end{eqnarray*}
By using (A.15) and (A.16), we see that each row $\beta_l(m+1)$
can be obtained from the row $\beta^0_l(m+1)$ and
$\Delta\beta_l(m+1)$ while the later two could be obtained from
$\beta_l(m)$ above. For example, using  $\beta_1(1)=\beta_0(1)=1$,
we have
\begin{eqnarray}
\beta^0_3(2)= \beta_1(1)\cdot (1+4) = 5, \nonumber\\
\beta^0_2(2)= \beta_1(1) \cdot 5 + \beta_0(1) \cdot 4  = 9, \nonumber\\
\Delta\beta_2(2)=\Delta\beta_3(2)=0\nonumber
\end{eqnarray}
etc. The values of the elements in the pyramid are
\begin{eqnarray*}
\begin{array}{rccccccccl}
\beta_l(1)&&&&1&1&&&&~~~m=1\\
&&&&&&&&&\\
\beta^0_l(2)&&&5&9&9&9&&&~~~\\
\Delta\beta_l(2)&&&0&0&0&0&&&~~~\\
\beta_l(2)&&&5&9&9&9&&&~~~m=2\\
&&&&&&&&&\\
\beta^0_l(3)&&35&89&134&170&170&170&&~~~\\
\Delta\beta_l(3)&&2&6&8&8&8&8&&~~~\\
\beta_l(3)&&37&95&142&178&178&178&&~~~m=3\\
&&&&&&&&&\\
\beta^0_l(4)&333&1093&2087&3155&4045&4757&4757&4757&~~~\\
\Delta\beta_l(4)&20&76&148&220&256&256&256&256&~~~\\
\beta_l(4)&353&1169&2235&3375&4301&5013&5013&5013&~~~m=4
~~~~~~~~~~~~~~~~~(A.19)\\
&&&\cdots&\cdots&&&&
\end{array}
\end{eqnarray*}
Correspondingly, the energies are
$$
\epsilon_2=\frac{9}{2^{6}},~~~\epsilon_3=\frac{178}{2^{10}}=\frac{89}{2^9},
~~~\epsilon_4=\frac{5013}{2^{14}},~~~{\rm etc.} \eqno(A.20)
$$

\vspace{1cm}

\noindent
{\bf Appendix B}\\

For the convenience of applying the coordinate system
$\{S,~\alpha\}$ defined in Eq.(2.10), the definition of some
quantities introduced in ref.[2] is given in the following. For
the new coordinate system
$$
(S,\alpha) = (S,\alpha_1({\bf q}),\alpha_2({\bf q}), \cdots,
\alpha_{N-1}({\bf q})), \eqno(B.1)
$$
each point ${\bf q}$ in the $N$-dimensional space will now be
designated by
$$
(S, \alpha_1, \alpha_2, \cdots, \alpha_{N-1}), \eqno(B.2)
$$
instead of $(q_1, q_2, q_3, \cdots, q_N)$. The corresponding line
element can be written as
$$
d\stackrel{\rightarrow}{{\bf q}} = \stackrel{\wedge}S h_S dS
+\sum_{j=1}^{N-1}\stackrel{\wedge}{\alpha}_j h_j d\alpha_j;
\eqno(B.3)
$$
the gradient is given by
$$
{\bf \nabla} = \stackrel{\wedge}S
\frac{1}{h_S}\frac{\partial}{\partial S}
+\sum_{j=1}^{N-1}\stackrel{\wedge}{\alpha}_j \frac{1}{h_j}
\frac{\partial}{\partial \alpha_j}. \eqno(B.4)
$$
The kinetic energy operator
$$
T=-\frac{1}{2} {\bf \nabla}^2
\eqno(B.5)
$$
can be decomposed into two parts:
$$
T = T_S + T_{\alpha},
\eqno(B.6)
$$
with
$$
T_S = -\frac{1}{2h_Sh_\alpha} \frac{\partial}{\partial S}
(\frac{h_{\alpha}}{h_S}\frac{\partial}{\partial S}),
\eqno(B.7)
$$
$$
T_{\alpha} = -\frac{1}{2h_Sh_\alpha} \sum_{j=1}^{N-1}
\frac{\partial}{\partial \alpha_j}
(\frac{h_Sh_{\alpha}}{h_j^2}\frac{\partial}{\partial \alpha_j}),
\eqno(B.8)
$$
in which
$$
h_{\alpha} = \prod_{j=1}^{N-1}h_j,
\eqno(B.9)
$$
and
$$
h_S^2 = [({\bf \nabla}S)^2]^{-1},~~~ h_1^2 = [({\bf
\nabla}\alpha_1)^2]^{-1},\cdots,~~~ h_j^2 = [({\bf
\nabla}\alpha_j)^2]^{-1},\cdots.
\eqno(B.10)
$$
The  volume element in the ${\bf q}$-space is now
$$
d^N {\bf q} = h_Sh_\alpha dSd\alpha
\eqno(B.11)
$$
with
$$
d\alpha = \prod_{j=1}^{N-1}d\alpha_j.
\eqno(B.12)
$$

\end{document}